\documentclass[12pt,preprint]{aastex}

\slugcomment{Accepted by the Astrophysical Journal}

\shorttitle{Crab Giant Pulses in 20--84~MHz}
\shortauthors{Ellingson et al.}

\begin{document}


\title{Observations of Crab Giant Pulses in 20--84~MHz using LWA1 }


%\author{S.W. Ellingson\altaffilmark{1}}
%\altaffiltext{1}{Bradley Dept.\ of Electrical \& Computer Eng., Virginia Tech, Blacksburg, VA 24060}
%\email{ellingson@vt.edu}
%%
%\author{G.B. Taylor\altaffilmark{2,3}}
%\altaffiltext{2}{Department of Physics and Astronomy, University of New Mexico, Albuquerque NM, 87131, USA}
%\altaffiltext{3}{Greg Taylor is also an Adjunct Astronomer at the National Radio Astronomy Observatory.}
%\email{gbtaylor@unm.edu}

\author{
S.W. Ellingson\altaffilmark{1},
T.E. Clarke\altaffilmark{2},
J. Craig\altaffilmark{3}, 
B.C. Hicks\altaffilmark{2},
T.J.W. Lazio\altaffilmark{4},
G.B. Taylor\altaffilmark{3,5},\\
T.L. Wilson\altaffilmark{2},
C.N. Wolfe\altaffilmark{1}
}
\altaffiltext{1}{Bradley Dept.\ of Electrical \& Computer Eng., Virginia Tech, Blacksburg, VA 24060.}
\altaffiltext{2}{US Naval Research Laboratory, Code 7213, Washington, DC 20375.}
\altaffiltext{3}{Department of Physics and Astronomy, University of New Mexico, Albuquerque NM, 87131.}
\altaffiltext{4}{Jet Propulsion Laboratory, California Institute of Technology, MS 138-308, 4800 Oak Grove
Dr., Pasadena, CA 91109.}
\altaffiltext{5}{Also Adjunct Astronomer at the National Radio Astronomy Observatory.}

%


%\altaffiltext{1}{Visiting Astronomer, Cerro Tololo Inter-American Observatory.
%CTIO is operated by AURA, Inc.\ under contract to the National Science
%\altaffiltext{2}{Society of Fellows, Harvard University.}
%\altaffiltext{3}{present address: Center for Astrophysics,
%    60 Garden Street, Cambridge, MA 02138}
%\altaffiltext{4}{Visiting Programmer, Space Telescope Science Institute}
%\altaffiltext{5}{Patron, Alonso's Bar and Grill}


\begin{abstract}
We report the detection and observed characteristics of giant pulses from the Crab Nebula pulsar (B0531+21) in four frequency bands covering 20--84~MHz using the recently-completed Long Wavelength Array Station~1 (LWA1) radio telescope.  In 10~hours of observations distributed over a 72~day period in Fall of 2012, 33 giant pulses having peak flux densities between $400$~Jy and $2000$~Jy were detected.  Twenty-two of these pulses were detected simultaneously in channels of 16 MHz bandwidth centered at 44~MHz, 60~MHz, and 76~MHz, including one pulse which was also detected in a channel centered at 28~MHz.  We quantify statistics of pulse amplitude and pulse shape characteristics, including pulse broadening.  Amplitude statistics are consistent with expectations based on extrapolations from previous work at higher and lower frequencies.  Pulse broadening is found to be relatively high, but not significantly greater than expected.  We present procedures that have been found to be effective for observing giant pulses in this frequency range.
\end{abstract}





\section{Introduction}

The Crab Nebula pulsar (B0531+21) intermittently produces single pulses having intensity orders of magnitude greater than those of the normal periodic emission \citep{sta68}. 
% Definition of a GP (how many times mean flux density) is...\\
% What percent of emission..\\
% Rate is about $\sim10-20 \times 10^{-3}$~s$^{-1}$ \citep{kar12}.  \citep{Lun95} gives rates at higher frequencies -- same?\\
% Emission mechanism unknown, so says \cite{mic12}, paragraph 2.\\
These giant pulses (GPs) consist of complex superpositions of nanosecond- and microsecond-scale impulses \citep{han03,cro10} and have been observed over many octaves of the radio spectrum \citep{sal99,cor04,pop06,bha07}. 
%Fluctuations in GP amplitudes appear to be associated with changes in the coherence of the radio emission \citep{lun95}. 
The details of the GP emission mechanism remain mysterious.  
Because the intrinsic duration of Crab GPs is much less than the pulse broadening due to the host nebula and the interstellar medium (ISM) along the line of sight, Crab GPs are also potentially of great value as a probe of the electron density of those regions. 

Studies of Crab GPs are often limited to frequencies above 100~MHz where the emission is relatively strong and/or existing large radio telescopes are available; see for example \cite{mof96}; 
\cite{cor04};
\cite{pop06};
\cite{bha07};
\cite{maj11};
\cite{kar12}; and
\cite{mic12}.  
Studies of Crab GPs below 100~MHz have historically been limited by the lack of suitable large telescopes and the difficulties imposed by strong and diverse radio frequency interference (RFI) at these frequencies. 
Observations of Crab GPs have been reported at 40 and 60~MHz by \cite{kuz02} and more recently at 23~MHz by \cite{pop06}.
A search of 14 hours of data collected in a 2.5~MHz bandwidth around 38 MHz by the 10-dipole Eight meter-wavelength Transient Array (ETA) experiment revealed 11 GP candidates,
%ed intensity ($\sim1$~kJy) obtained from log-linear extrapolation between published detections at 23~MHz and 200~MHz, 
but these proved difficult to confirm due to the low sensitivity ($\sim4\sigma$ for the strongest pulse candidate of $\sim1$~kJy) and difficulties with RFI \citep{des09}.
Detection of Crab GPs by LOFAR in the frequency ranges 32--80~MHz and 139--187~MHz has been reported but further details are not provided \citep{sta11}. 

In this paper we report the detection and characterization of Crab GPs in the frequency range 20--84~MHz by the recently-completed Long Wavelength Array Station~1 (LWA1) radio telescope.  LWA1 consists of a pseudorandom array of 256~dual-polarized broadband dipoles within a 110~m $\times$ 100~m elliptical footprint,\footnote{Not including a small number of ``outrigger'' dipoles which lie $>300$~m from the center of the array and are used for array calibration and power spectrum measurement.} and is capable of Nyquist-rate acquisition of beam outputs covering its entire tuning range of 10-88~MHz simultaneously.
% which is capable of Nyquist-sampled observations of it's entire 10--88~MHz tuning range simultaneously with nearly the best possible (that is, Galactic noise-limited) sensitivity. 
This facilitates high-signal-to-noise ratio detection of Crab GPs
%, which are relatively weak in this frequency range, 
at a rate of several per hour; also, the large instantaneous bandwidth provides the ability to study variations in flux density and pulse broadening with frequency on a pulse-by-pulse basis.
The large bandwidth also facilitates improved methods for the recognition and mitigation of the effects of RFI.
% No distinction between main pulse and interpulse.\\

This paper is organized as follows:  Section~\ref{sIDA} describes LWA1 and the methods used for data collection, data reduction, identification of Crab GPs, pulse characterization, and flux density calibration. Section~\ref{sResults} summarizes the findings of the study including statistics of pulse amplitude and pulse broadening. Section~\ref{sDisc} includes a brief description of ongoing and planned follow-up work.

\section{\label{sIDA}Instrumentation \& Data Analysis}

\subsection{LWA1}

LWA1 is collocated with the Very Large Array (VLA; $107.63^{\circ}$~W, $34.07^{\circ}$~N) in central New Mexico.  The telescope consists of a primary array of 256 pairs of dipole-type antennas whose outputs are individually digitized and formed into beams.  A detailed technical description of the instrument, including performance characterization, can be found in \cite{ell13}.  
LWA1 provides 4 independently-steerable dual-polarization beams.
Each beam provides two separate passbands with independently-selectable center frequencies.
Each passband consists of Nyquist-rate time-domain output at sample rates up 19.6 million samples per second (MSPS) per passband with $\sim17$~MHz usable bandwidth, in 4-bits real (``I'') plus 4-bits imaginary (``Q'') format.  
The beam main lobe width varies with frequency and pointing direction, but ranges from about $23^{\circ}$ full width at half-maximum (FWHM) at $20$~MHz to about $3^{\circ}$ FWHM at $84$~MHz for observations of the Crab Pulsar within about 1 hour of upper culmination.  The system equivalent flux density (SEFD) varies as a function of frequency, pointing relative to zenith, and (because the system temperature is strongly dominated by Galactic noise) pointing in celestial coordinates.\footnote{SEFD is defined here as the flux density of an unresolved source in the main lobe which results in a doubling of the measured beam output power relative to the value in the absence of the source.}
SEFD estimates for the flux density calibration specific to the present work are developed in Section~\ref{sFC}; however SEFD is typically found to be in the range 6--17~kJy for high-elevation pointings over most of the frequency range.\footnote{The counterintuitive fact that SEFD is approximately constant over a $\sim4:1$ frequency range can be understood by noting that the collecting area of a dipole decreases as the square of frequency, whereas the Galactic noise-dominated antenna temperature decreases with frequency at about the same rate.}  This corresponds to a $5\sigma$ sensitivity $< 100$~Jy assuming both polarizations, 17~MHz bandwidth, and 25~ms integration. 

A description of LWA1 early science results (including first pulsar observations) can be found in \cite{tay12}. To date LWA1 has detected 13 known pulsars and has several on-going programs dedicated to survey, monitoring, and characterization of pulsars \citep{sto13}. 

\subsection{\label{sDC}Data Collection}

Data for the present work were collected in 10 1-hr observations in Fall 2012, each centered on the transit of the Crab Pulsar, as detailed in Table~\ref{tObs}.  In each observation, two beams were used together to track the pulsar (i.e., both beams pointing in precisely the same direction).  The center frequencies of the passbands of the first beam were 76~MHz and 60~MHz, whereas the center frequencies for the second beam were 44~MHz and 28~MHz.  Since the usable bandwidth is $> 16$~MHz per passband, this scheme yields contiguous frequency coverage from 20~MHz to 84~MHz.  
%One 2-hour observation from UTC Mar 3, 2012, using only 1 60~MHz tuning from one beam, was also included.  
%The total observing ``wall time'' was 7 hours, and 
The total volume of data collected in this fashion was $\sim5$~TB; relatively large since the data are recorded as raw voltage streams without channelization or other ``on the fly'' processing.  This format provides the maximum flexibility for both initial processing and subsequent reprocessing of the data at later time using other search parameters or techniques. 

Although the instrument is capable of tracking sources to very low elevation angles, we have restricted the observations to 1 hour each day around upper culmination in order to (1) limit the extent to which change in the beam characteristics over the track can manifest as variations in estimated flux density, (2) distribute the observations more evenly over the expected range of source conditions, and (3) reduce the tempo of data acquisition to a manageable pace.  Concerning (2), a parameter of interest in the present work that is known to vary significantly over timescales of weeks to months is pulse broadening \citep{kuz08}; in fact, such variations can occur even on timescales of minutes \citep{kar10}.

Because this study involved the use of a new instrument and new data analysis procedures, we 
%were concerned about the possibility of a problems with either the instrument or new data analysis software interfering with our ability to detect Crab GPs *FIXME(previous sentence is awkward)*.  For this reason, some observing runs included 
performed
simultaneous observations of the bright pulsar B0329+54 using the remaining two beams, using the same synchronized-tracking four-passband scheme employed for Crab GP observing.  This allowed an independent check of the instrument and data reduction procedures using a well-known object, as well as providing an ``off pointing'' (separated by $\sim2$~h in right ascension) for confirmation of Crab GP detections as well as for identification of RFI through anticoincidence techniques.  

\subsection{\label{sDR}Data Reduction}

The data were analyzed using custom incoherent dedispersion software developed by one of us (S.E.), designed specifically for rapid analysis of periodic and single-pulse emission of dispersed astrophysical signals in raw LWA1 beam data.  The steps in data reduction for one passband of one beam were as follows:  First, any samples having $\sqrt{I^2+Q^2} \ge 7$ (that is, the magnitude of the complex-valued sample equal to or greater than maximum encodable real or imaginary magnitude) were assumed to be due to RFI and set to zero; typically 0.1\% to 1\% of samples were zeroed in this process. Next, data were partitioned into blocks of length $L$ (center frequency-dependent) and divided into channels using the fast Fourier transform (FFT).  
%$L$ was 16384, 16384, 65536, and 131072 for the 76, 60, 44, and 28~MHz passbands, respectively.  The associated channel bandwidths $\times$ integration times were 1196~Hz $\times$ 836~$\mu$s, 1196~Hz $\times$ 836~$\mu$s, 299~Hz $\times$ 3.34~ms, and 150~Hz $\times$ 6.69~ms, respectively, and 
Values of $L$ and the associated time and frequency resolutions are shown in Table~\ref{tFFT}.
These values were chosen to make the residual dispersion negligible relative to the 
%dedispersed pulse width.  
width of the scatter-broadened pulse after incoherent dedispersion.
The raw linear polarizations were incoherently combined at this point.  Time-frequency pixels associated with the outer 10\% of frequency channels were discarded and the inner 90\% were ``flattened''; that is, variations in frequency and time (attributable to the instrument or to ionospheric effects) within the time span of the spectrogram (typically 10--50~s) were removed. 
Finally, all pixels in frequency bands containing persistent RFI, or RFI which was intermittent but frequently observed, were discarded; additional details about this procedure are given in Section~\ref{sIGP}.  

In addition to the amplitude variation, the ionosphere imparts both refraction and dispersion to received signals.  For the purposes of the present work, these additional effects are negligible and may be safely ignored:  The worst-case ionospheric refraction is on the order of arcminutes (see e.g. \cite{k+07}) and thus is a tiny fraction of the narrowest beam width in the present work.  The worst-case ionospheric contribution to the total dispersive delay over 20--84~MHz is $\ll 1$~ms \citep{rjt01}, corresponding to an apparent error in dispersion measure (DM) $< 10^{-5}$~pc~cm$^{-3}$.  This is at least two orders of magnitude less than the error in assumed DM used for dedispersion, which is described below and shown to be insignificant.      

Following RFI mitigation, the result was incoherently dedispersed assuming DM $=$ 56.791 pc~cm$^{-3}$ \citep{man05}.  Although coherent dedispersion would be preferable from a sensitivity perspective, the residual dispersion after application of incoherent dedispersion at the time-frequency resolutions identified above is less than than the observed width of the scatter-broadened pulses by more than two orders of magnitude; thus the penalty in the present work is negligible.  In addition to having a large advantage in terms of computation burden, the frequency-domain interference mitigation procedure described below (Section~\ref{sIGP}) is easily implemented in the incoherent dedispersion framework, but much less straightforward to implement as part of a coherent dedispersion scheme.   The dedispersed time series was averaged to obtain an integration time and effective time resolution of approximately 25~ms for the 76~MHz and 60~MHz passbands, and either 33~ms or 50~ms for the 44~MHz and 28~MHz passbands.  

Note that there is no subsequent adjustment to the assumed DM of 56.791~pc~cm$^{-3}$ in our processing.  In particular, we did not perform a search over DM.  As explained in Section~\ref{sPB}, refinement of the DM used for dedispersion proves to be unnecessary because the resulting bias in estimated pulse parameters turns out to be negligible.
%compared to the values of the parameters themselves.

\begin{figure}
%\epsscale{.80}
\plotone{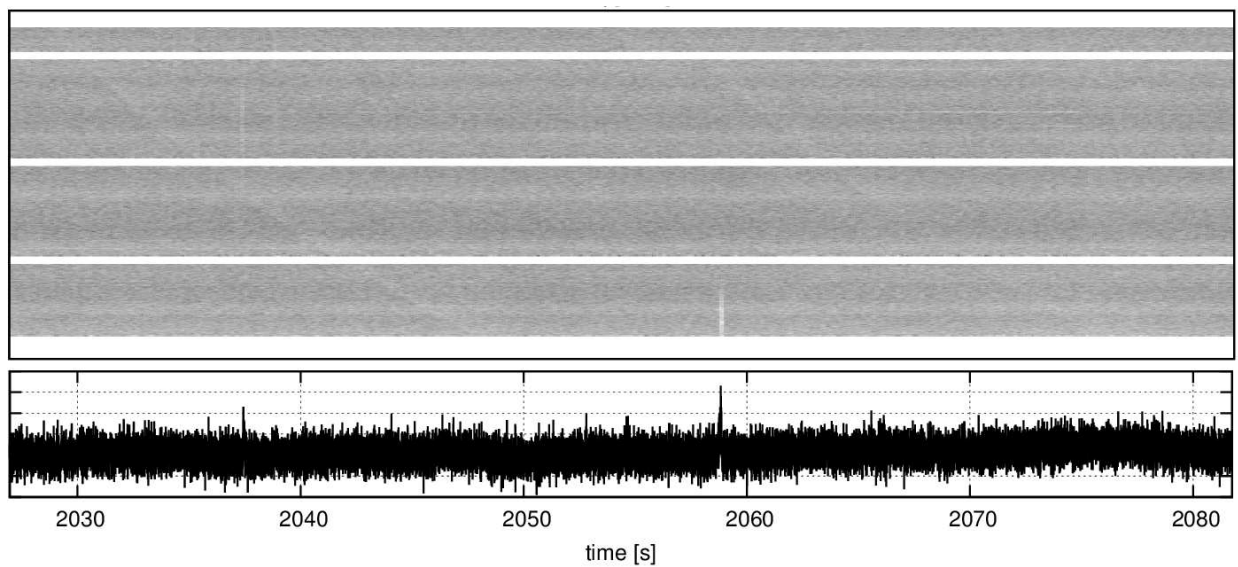}
%\vspace{1in}
\caption{
A pair of Crab GPs displayed in joint time-frequency (``spectrogram'') representation (top) and total power (bottom).
The vertical axis of the spectrogram is a 19.6~MHz span centered at 76~MHz, with frequency increasing toward the top.
Pulses begin at 2029~s and 2057~s at the top of the spectrogram, and ``chirp'' toward the bottom right.
Horizontal white bands are excised frequency channels, including the outer 10\% of the Nyquist bandwidth and (from bottom to top)
NTSC channel 4 audio, NTSC channel 5 video, and NTSC channel 6 video.
The vertical streaks at 2037~s and 2057~s are ``flares'' from ATSC channels 6 and 4, respectively.
The resolution of the spectrogram is approximately 100~kHz $\times$ 5~ms.  The resolution of the time series is $836~\mu$s.  
\label{fNicePulse}}
\end{figure}

The products saved from this procedure were (1) the dedispersed and averaged time series, and (2) the associated spectrogram, which was further averaged to a time-frequency resolution of approximately 5~kHz $\times$ 100~ms.  A representative spectrogram is shown in Figure~\ref{fNicePulse}.  The spectrograms play an important role in data interpretation, as explained below.

In observing sessions where we simultaneously observed B0329+54, we repeated the above procedure in all details, except for the DM (26.833~pc~cm$^{-3}$) and we took the additional step of folding the resulting time series at the known period ($\approx0.714$~s) \citep{man05}.  In each case we detected the pulsar in all four passbands with high signal-to-noise ratio, giving confidence in the new instrumentation and data analysis procedures.  Furthermore we confirmed that GPs detected in the beams pointing to B0531+21 were absent from the beams pointing to B0329+54, as expected.  

\subsection{\label{sIGP}Identification of Crab GPs}

Traditional methods for identification of GPs involve matched filtering of the time series, typically with iteration over DM in order to confirm that detected pulses are in fact Crab GPs.  A straightforward application of this procedure is not robust for the present work, for several reasons.  First, ionospheric scintillation results in variations that can vary by as much as 10\% of the system temperature with characteristic time scales on the order of minutes.\footnote{For examples, see Figures 10--11 of \cite{ell13}.}  This frequently results in large variations in the noise baseline of the time series, and occurs on time scales roughly equal to or less than the time it takes for the ``chirp'' waveform of the dispersed pulse to traverse the passband.  Furthermore, GPs in this frequency range remain scatter-broadened to durations on scales of 100--1000~ms after dedispersion.  This tends to make matched filter detection procedures that assume canonical pulse shapes with time-invariant noise baselines unreliable and frustrating to use, especially in the presence of impulsive RFI.  

Furthermore, impulsive RFI is ubiquitous in this frequency range.   In particular, we have found that signals from terrestrial television transmitters in the frequency range 54--87 MHz are a serious problem.  Transmissions in both the modern digital ``ATSC'' format currently in widespread use in the U.S., as well as transmissions in the legacy analog ``NTSC'' format, are routinely found in LWA1 data.  ATSC signals are recognizable as emissions having 6 MHz instantaneous bandwidth, whereas NTSC signals are recognizable as relatively narrowband video and audio carriers separated by 4.5~MHz.  With the exception of a persistent NTSC video carrier seen at 55.25~MHz (associated with a station in Mexico), these emissions are usually not seen as continuous transmissions, but rather in bursts with durations on the order of milliseconds to seconds.  These bursts are believed to be due to reflections from ionized meteor trails and low Earth orbiting satellites, as opposed to reflection from aircraft or the ionosphere.\footnote{This is based on a preliminary and ongoing analysis of LWA1 data by J. Helmboldt of the U.S. Naval Research Laboratory.}  
%Figure~\ref{fRFI} shows an example.  
These emissions have two particularly irritating features from the perspective of searching for single dispersed pulses:  (1) They frequently exhibit the same ``reverse ramp'' waveform expected of scatter-broadened pulses, and with similar durations, and (2) Because they are narrowband relative to the overall bandwidth, they are often not significantly suppressed by the process of dedispersion -- they are simply translated in time and (in the case of ATSC flares) also ``smeared'' in time.  

Excluding bandwidths of 1.0 and 0.5~MHz around frequencies of NTSC video and audio carriers (respectively) dramatically improves -- but does not eliminate -- the NTSC problem.  Specifically, strong NTSC flares occasionally ``overrun'' the excluded frequency ranges. ATSC flares cannot be effectively mitigated by frequency blanking without removing the entire 6~MHz bandwidth of the emission.  Reducing the effect of NTSC and ATSC flares to a negligible level by frequency-domain blanking requires eliminating an unacceptable amount of otherwise-usable bandwidth.  Because the flares occur with relatively small overall duty cycle, an automated joint time-frequency blanking scheme probably would be acceptable; however, this requires significant software development and has not yet been implemented.

At present, we find that the most practical and robust method for detection of Crab GPs in 
%the presence of impulsive RFI and other impairments 
our frequency range
is as follows:  First, we manually search the spectrograms for each passband for the characteristic ``chirp'' of a Crab GP which is visible over most of the passband.  Figure~\ref{fNicePulse} shows an example.    We then confirm the presence of the distinctive broadened pulse waveform at the appropriate time in the dedispersed time series.  Typically -- but not always -- pulses are first detected in the 76~MHz or 60~MHz passbands; this is because the pulse broadening is less severe, and because pulse flux density in this range of frequencies generally increases with increasing frequency \citep{pop06,kar12}.  The dedispersed time series for the other passbands is then also checked, accounting for the expected dispersion delay which is tens to hundreds of seconds due to the large fractional bandwidth between passbands.  
The first 10 or so GPs that we detected using this approach were subsequently independently confirmed by repeating the procedure 
while sweeping in DM.
%for lower and higher trial DMs and noting non-detection these DMs.  
We soon realized that that this computationally-expensive procedure was not necessary for the present work, since simpler procedure of detecting pulses through spectrograms and/or appearance in multiple passbands with the expected dispersion delay yielded a false alarm rate of zero, whereas the results of DM sweeps were frequently difficult to interpret due to the effects of the ionosphere and impulsive RFI, as discussed above.
%When a GP is detected in any of the four tunings using the above procedure, we search for the pulse at the expected time in the dispersed time series for the other three tunings and accept the detection if the apparent S/N is greater than 10.  
%Most of the detections reported here are made initially in the 76 or 60~MHz tunings, and are subsequently found in the 2 or 3 highest-frequency tunings using this technique.  
%The sole 28~MHz band detection reported here was found through a particularly strong ``primary'' detection in the 44~MHz tuning.

A possible disadvantage of this detection procedure is somewhat reduced sensitivity.
Manual inspection of dedispersed time series reveals many pulse candidates which are quite likely to be Crab GPs, but which are not visible in spectrograms and/or ambiguous or non-detections in the dedispersed time series for other passbands. 
We suspect refinements to the above procedures can increase the rate at which GPs are detected by a factor of 2--3 at least.  For this initial work, however, we prefer a simpler and more conservative approach.
Furthermore (as discussed in Section~\ref{sPC}), we shall apply an even more stringent criterion of S/N $> 4.4\sigma$ for the peak flux density.  Several candidate GPs detected using the above procedure failed this criterion, indicating that the sensitivity penalty is actually not unusually severe in comparison to search procedures (including fully-automated techniques) employed by others; e.g. \cite{bha07} and \cite{kar12}.

Finally, we note that frequencies below 54~MHz down to the lowest frequency considered in this study ($\sim20$~MHz) exhibit relatively sparse and low levels of anthropogenic RFI during local hours of darkness, when all of our observations were conducted.  We found that no frequency blanking was required for the 28~MHz passband, and only 0.2~MHz was blanked in the 44~MHz passband (to remove a persistent tone -- possibly locally generated -- at 48.0~MHz).  Ironically, ionospheric scintillation is significantly worse at the lower frequencies; in some cases the resulting variations in the noise baseline were large enough to make characterization of the detected pulses intractable, and so these pulses were excluded.  
%Nevertheless it is believed that it is the intrinsically low flux density of pulses in this frequency range, as opposed to the disposal of pulses distorted by baseline variations, that is primarily responsible for the relatively low rate of detections in the 28~MHz and 44~MHz passbands.
Nevertheless it is believed that it is the low flux density of pulses relative to the system sensitivity, as opposed to the disposal of pulses distorted by baseline variations, that is primarily responsible for the relatively low rate of detections in the 28~MHz and 44~MHz passbands.

\subsection{\label{sPC}Pulse Characterization}

Below 84~MHz, the duration of Crab GPs is broadened by at least three orders of magnitude by scattering within the host nebula and along the line-of-sight through the ISM.  Thus it is an excellent approximation to assume that 
%the intrinsic pulse shape is an impulse having bandwith $\gg$ the receive bandwidth, and therefore 
the shape of the received pulse is simply a scaled replica of the scattering impulse response $g(t)$.  Appropriate pulse shapes are given by \cite{wil72} as follows:
\begin{equation}
g(t) ~\propto~ \left( 1 / \tau_d \right) \exp\left( -t / \tau_d \right) u(t)
\label{eg1}
\end{equation}
corresponding to scattering by a thin slab of material of infinite extent, and
\begin{equation}
g(t) ~\propto~ \left( \pi^5 \tau_d^3 / 4 t^5 \right) \exp\left( -\pi^2 \tau_d / 4t \right) u(t)
\end{equation}
corresponding to scattering by material which is uniformly-distributed along the line of sight. In both forms $u(t)$ is the unit step function ($u(t)=0$ for $t<0$, $u(t)=1$ for $t\ge0$; $t=0$ taken to be the pulse start time) and $\tau_d$ is referred to as the {\it characteristic broadening time}.  The dependence of $\tau_d$ on frequency in the first form is determined by the spatial distribution of electron density.  When this distribution follows the Kolomogorov spectrum, $\tau_d \propto \nu^{-4.4}$, and early measurements seemed to indicate that this was the case \citep{isa77,sal99}.  Subsequent observations indicate a considerably more shallow dependence, with exponents in the range $-3.5$ to $-3.2$ \citep{pop06,bha07,kar12}.  \cite{kar12} note that pulses observed in the 116--174~MHz band are not well described by Equation~\ref{eg1}, but obtain good fits to the modified form
\begin{equation}
g(t) ~\propto~ t^{\beta} \exp\left( -t / \tau_d \right) u(t)
\label{eMP}
\end{equation}
which includes a rounded leading edge with rise time determined by the parameter $\beta$.  We also find in the present work that the above form provides the best overall fit to the observed pulses, whereas the other forms generally do not; thus we have adopted this model.  As in \cite{kar12}, we constrain $\beta$ to be in the range 0--1 and find that this range seems to accommodate all pulses encountered.

\begin{figure}
%\epsscale{.80}
\plotone{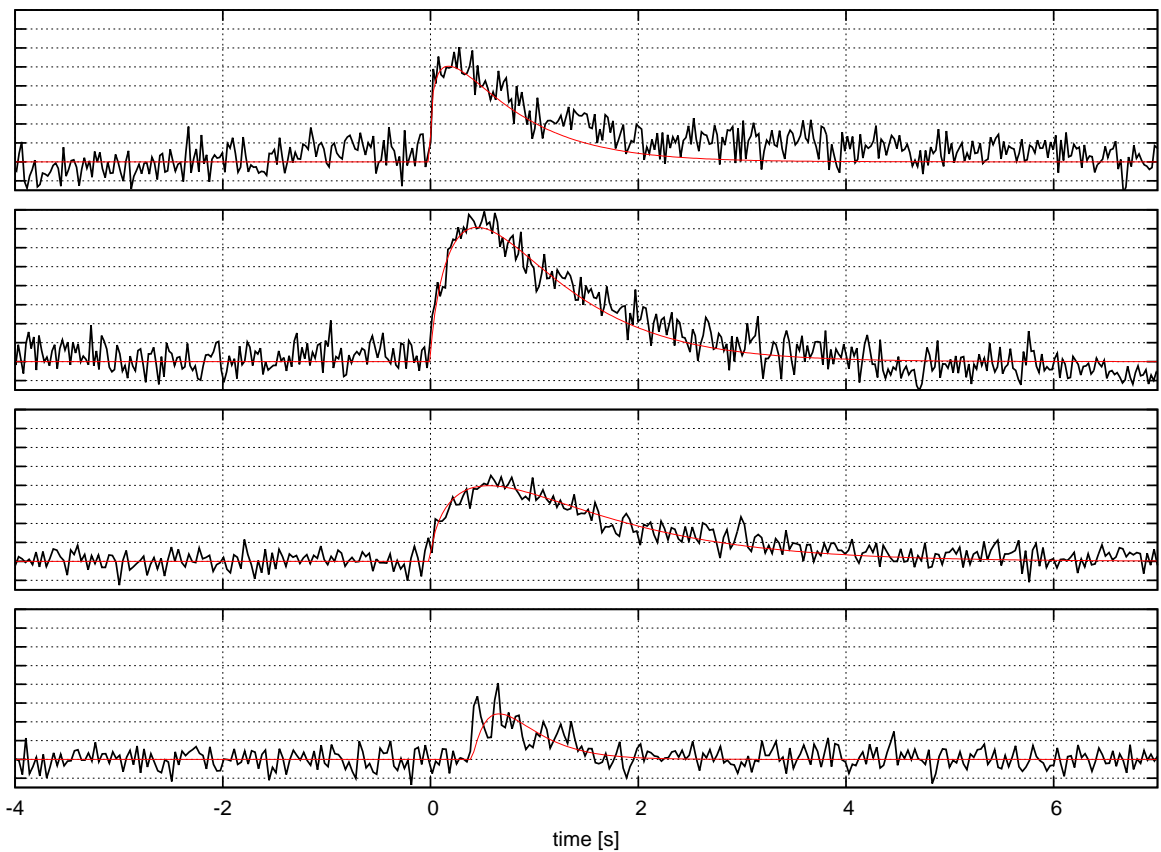} %\plotone{19.eps}
%\vspace{1in}
\caption{
A pulse observed simultaneously in the 76, 60, 44, and 28~MHz passbands; top to bottom, respectively.  Dispersion and dispersive delay between passbands has been removed, leaving only pulse broadening.  The best-fit model pulses are also shown (note that most of the rising edge of the 28 MHz pulse is not visible due to inadequate sensitivity).  The vertical scales are unmodified from the raw data, and thus include instrumental variations in gain and sensitivity between passbands.  Date of observation: MJD 56291.
%used to determine the pulse broadening time $\tau_d$ are also shown; in this case $\tau_d$ is found to be 0.3, 0.8, and 1.0~s, respectively.  
%Amplitudes in each plot have been normalized to the same maximum value. 
\label{fPulseFit}}
\end{figure}

Thus, each pulse in the present work is characterized by $\tau_d$, $\beta$, and amplitude. These are determined as follows:  First, we fit a second-order polynomial to the time preceding the start of the pulse (i.e., $t<0$) in the dedispersed time series.  The baseline variation is removed, and then an automated brute-force fit in three dimensions ($\tau_d$, $\beta$, and amplitude) is performed to find the least-squares fit to the data.  Figure~\ref{fPulseFit} shows examples.

For verification, the model pulse is then subtracted from the data to confirm that the resulting time series is flat and free from artifacts.  This turns out to be particularly important in our frequency range, since detectable pulses sometimes overlap each other and thus bias the estimation of parameters.  In the present work, two otherwise-qualified pulses were excluded from consideration (and thus do not appear in Table~\ref{tObs}) for this reason.    

The amplitude of the pulse is characterized in two ways.  First, the pulse peak S/N is calculated as the ratio of the peak value of the model pulse to the pre-pulse noise power $N$.  This is the value used in the $4.4\sigma$ detection criterion, described above.  Because the integration time $\Delta t$ (25--50~ms) is at least an order of magnitude less than the apparent width of the pulse, the {\it effective} S/N is much greater, and is given by
\begin{equation}
\left[\left( E_g / W \right) / N \right] \sqrt{\Delta t/\ W_{int} }
\label{eSNI} 
\end{equation}
where $E_g$ is the pulse energy, $W$ is the ``effective width'' of the pulse, and $W_{int}$ is the intrinsic pulse width, assumed to be $300~\mu s$ \citep{bha07}.   The effective width $W$ is the width of a rectangular pulse having the same peak amplitude and energy as the model pulse.  For the form given in Equation~\ref{eMP}, $W \approx \tau_d \exp{(\sqrt{\beta})}$.  Note that the effective S/N can only be realized {\it for detection} by either optimal pulse-matched filtering or by deconvolution of $g(t)$ from the intrinsic pulse shape \citep{bha03}.  Thus the estimated value obtained using the above procedure is used only for flux density calibration (and not detection), as described in the next section.
%The second factor in Equation~\ref{eSNI} accounts for an increase in S/N that can be achieved only by deconvolving $g(t)$ from the intrinsic pulse, as described in \cite{bha04}.

\subsection{\label{sFC}Flux Density Calibration}

Flux density calibration is challenging for array radio telescopes operating at low frequencies.  The primary difficulty is that beam characteristics vary as a function of frequency and pointing relative to zenith, pointing relative to the celestial sphere (because the system temperature may be dominated by sky noise), and source magnitudes can be modulated by as much as 10\% due ionospheric scintillation.  See \cite{ell13} for a discussion of these factors as they apply to LWA1.   

For the present work we used the transit drift scan SEFD measurement procedure described in Appendix~B of \cite{ell13}.  
In this procedure, the data are collected in exactly the same manner described in Section~\ref{sDC} with the exception that the beams are fixed, pointing at the point of upper culmination of the pulsar, and the total duration of the observation is 2~hours. 
 The data are reduced to full-bandwidth total power (single channel) time series.  
The transit of Tau~A (the unresolved radio source which includes the Crab Pulsar and the surrounding nebula) is clearly visible as a broad peak in the time series as the source moves through the main lobe.  When the source is outside the main lobe, the time series is thus a representation of the system temperature of the instrument in the vicinity of, but excluding Tau~A.  The peak of the time series can thus be interpreted as the sum of the SEFD and the flux density of Tau~A.  Assuming the flux density of Tau~A is known, the SEFD can in principle be calculated from the ratio of the peak of the time series to the level of the flat, off-peak portion of the time series.  In practice, we estimate and subtract a model of the beamshape, and use the level of the resulting noise baseline at the time of transit as the ``source turned off'' value; this reduces the error associated with the gradual change in SEFD over time.

In the present work we ignore the variation in beam characteristics with source position (in particular, source elevation). Since each observation is limited to 1~hour around upper culmination, this is variation is believed to be negligible.  For example: Assuming a $\cos^2Z$ dependence of beam directivity with zenith angle $Z$, the associated change of directivity over a 1~hour observation is less than 2\%.

The flux calibration procedure requires an accurate value for the flux density of Tau~A over the range of frequencies considered.  Our model for the flux density of Tau~A is
\begin{equation}
\left( 1944~\mbox{Jy} \right) \left( \nu / 76~\mbox{MHz} \right)^{-0.27}
\end{equation}
The spectral index in this model is the one derived in \cite{bar77}.  The reference flux density of 1944~Jy at 76~MHz is the average of values extrapolated from measurements at 22.25~MHz and 81.5~MHz by \cite{rog69} and \cite{par68}, respectively,\footnote{These measurements are also reported in \cite{bar77}} using this spectral index. This model gives values which fall within the error bars of the 22.25~MHz and 81.5~MHz measurements ($\pm6$\% and $\pm4$\%, respectively) and turns out to be essentially the same as model used in recent work on Crab GPs at low frequencies (e.g., \cite{bha07,kar12}), attributed to \cite{bie97}, which uses the same spectral index but gives a 76~MHz flux density of 1915~Jy.  Uncertainty in the reference flux density ($\pm5$\%) and the spectral index ($\pm0.04$), together result in an uncertainty of about $\pm10$\% at 20~MHz relative to the 84~MHz value.  In practice the model may also be biased at the lowest frequencies by factors not taken into account; e.g., turnover in the spectrum of the nebula.
%
% OLD VERSION: 
%This model was obtained from the spectral index derived in \cite{bar77}, which is then used to extract a best-fit 76~MHz flux density from measurements at 22.5~MHz and 81.5~MHz by \cite{rog69} and \cite{par68}, respectively.\footnote{These measurements are also reported in \cite{bar77}} 
%*FIXME(not sure this makes sense -- check)*
%It should be noted that this model is essentially the same as model used in recent work on Crab GPs at low frequencies (e.g., \cite{bha07,kar12}), attributed to \cite{bie97}), which uses the same spectral index but gives a 76~MHz flux density of 1915~Jy.  Uncertainty in the reference flux density ($\sim5$\%) or the spectral index ($\pm0.04$), together result in an uncertainty $\sim20$\% at 20~MHz relative to the 76~MHz value.  In practice the model may also be biased at the lowest frequencies by factors not taken into account; e.g., turnover in the spectrum of the nebula.

Using the above approach we obtain SEFD of 13.7~kJy, 10.6~kJy, 10.3~kJy, and 11.5~kJy for the 76, 60, 44, and 28~MHz passbands, respectively. The 76~MHz value is approximately twice the 74~MHz value reported in \cite{ell13}, which was obtained using essentially the same technique.  The difference is due to a recalibration of the array beamforming delays which occurred before the observations reported here were started, and appears to have degraded the beamforming performance.\footnote{A recalibration of the array in March 2013 has restored the original sensitivity.}
%, and has been noted in other unrelated recent observations.  
No earlier measurements of the same pointing are available for the other passbands; however the values reported above fall within the range 6--17~kJy that we normally expect for high-elevation sources.  The clear presence of a peak at the expected time in the flux density calibration drift scans confirms pointing was working properly for all passbands during the observations.

%another Tau~A derived SEFD reported in \citep{ell12} as well as the SEFD estimated from an electromagnetic simulation of LWA1 assuming uniform sky brightness, described in \citep{ell11} and also shown in \citep{ell12}.  Values for 44~MHz and 28~MHz are bound to be less accurate as the small errors in the flux and spectral index at the reference frequency has a larger effect; and the spectrum of Tau~A is approaching it's turnover frequency, limiting the accuracy of log-linear extrapolation.

From the above SEFD estimates we find the $4.4\sigma$ sensitivities (in terms of the effective S/N as described in the previous section, and not in terms of the peak waveform S/N) are
656, 515, 444, and 493~Jy for the 76, 60, 44, and 28~MHz passbands, respectively.
These figures take into account frequency-domain blanking for RFI purposes, resulting in bandwidths of 
14.07, 13.64, 17.44, and 17.64~MHz for the 76, 60, 44, and 28~MHz passbands, respectively.

\section{\label{sResults}Results}

The observing sessions and the resulting detections are summarized in Table~\ref{tObs}.  
A total of 33 Crab GPs were identified in 10 hours of observation, distributed over 72 days during Fall 2012.
One pulse was detected simultaneously in all 4 passbands, 21 pulses were detected simultaneously in 3 passbands, and an additional 11 pulses were detected simultaneously in 2 passbands.
All but a handful of pulses detected passed the $4.4\sigma$ peak S/N criterion.

\subsection{Pulse Amplitudes}

\begin{figure}
%\epsscale{.80}
\plotone{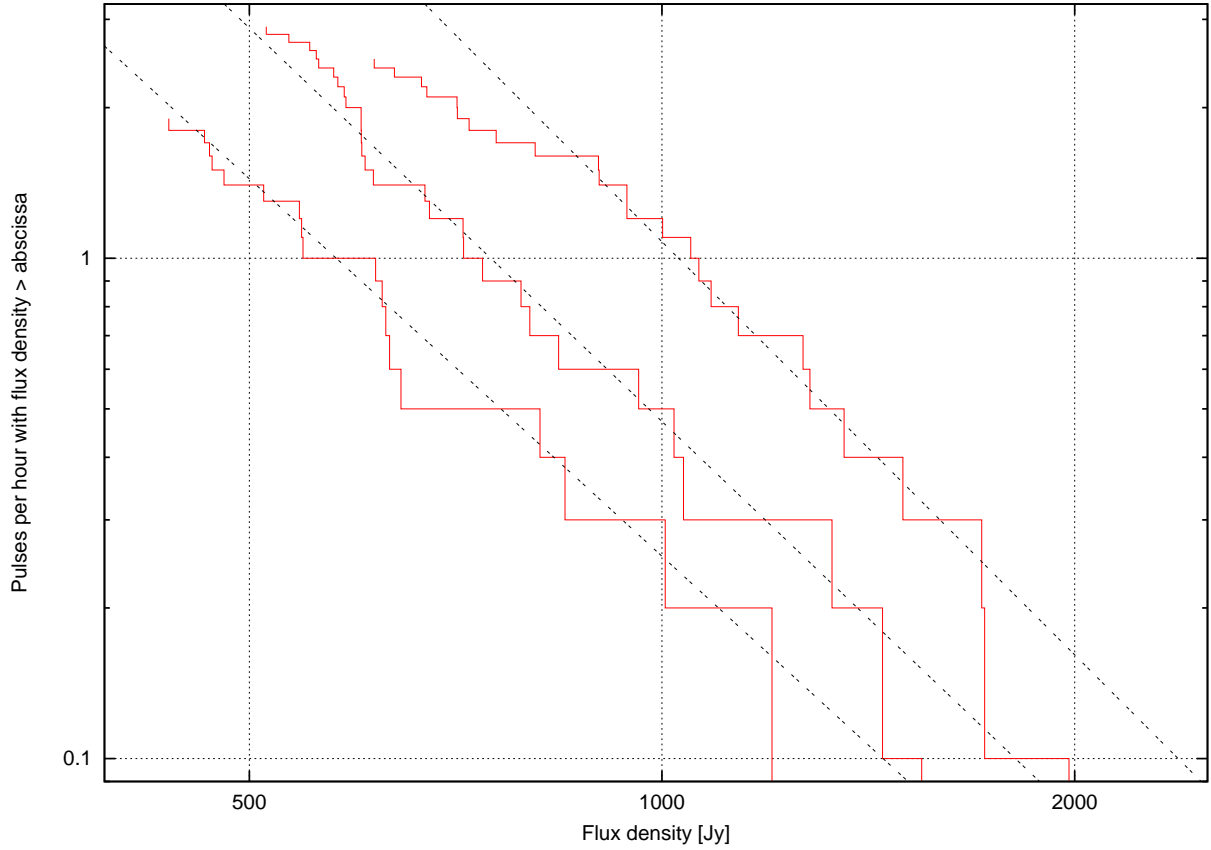}
%\vspace{1in}
\caption{
Detection rates expressed as number of pulses detected with flux density $>$ the abscissa.  Rates are expressed as the total number of detections divided by the number of hours of observation.  The curves correspond to the (left to right) 44, 60, and 76~MHz passbands.  (No data are shown for the 28~MHz passband since there was only one detection.) Also shown is a power law fit to the 18 largest values in each passband. 
\label{fCDFs}}
\end{figure}

Traditionally, statistical distributions of pulse amplitudes are expressed as power law fits to the tail distributions of the corresponding probability distribution functions (PDFs, or ``histograms''); see for example \cite{arg72}; 
\cite{maj11}; and\\ 
\cite{kar12}.  
This form of analysis is not meaningful for the present work due to the relatively small number of pulses detected.  Instead, Figure~\ref{fCDFs} summarizes detections as amplitude cumulative distribution functions (CDFs); i.e., expected rates of detections for pulses $\ge$ a specified flux density.  The CDF is the integral of the PDF; however the tail distributions of the two functions are not related in a straightforward way.  Nevertheless for the purposes of pulse rate prediction in this and future work, a power law fit to CDFs shown in Figure~\ref{fCDFs} was done, with the following results:
\begin{equation}
N_{76} = \left( 1.08~\mbox{h}^{-1} \right) \left( S / \mbox{kJy} \right)^{-2.74}
\label{eN76}
\end{equation}
\begin{equation}
N_{60} = \left( 0.47~\mbox{h}^{-1} \right) \left( S / \mbox{kJy} \right)^{-2.61}
\label{eN60}
\end{equation}
\begin{equation}
N_{44} = \left( 0.25~\mbox{h}^{-1} \right) \left( S / \mbox{kJy} \right)^{-2.51}
\label{eN44}
\end{equation}
Where $N_{\nu}$ is the number of pulses expected with flux density $\ge S$ at $\nu$~MHz. (No such characterization is possible for the 28~MHz passband, since one pulse was detected.)  Note that the distributions exhibit very similar dependence on frequency, and may be steepening with increasing frequency.  Although these values cannot be directly compared to analogous characterizations of the PDFs, we note that the same trend is seen in past work using the PDFs. 
%
%  76 MHz: (1.831257e+08 /hr)*S^(-2.743226)
%  60 MHz: (3.240358e+07 /hr)*S^(-2.612349)
%  44 MHz: (8.383371e+06 /hr)*S^(-2.506377)

\begin{figure}
%\epsscale{.80}
\plotone{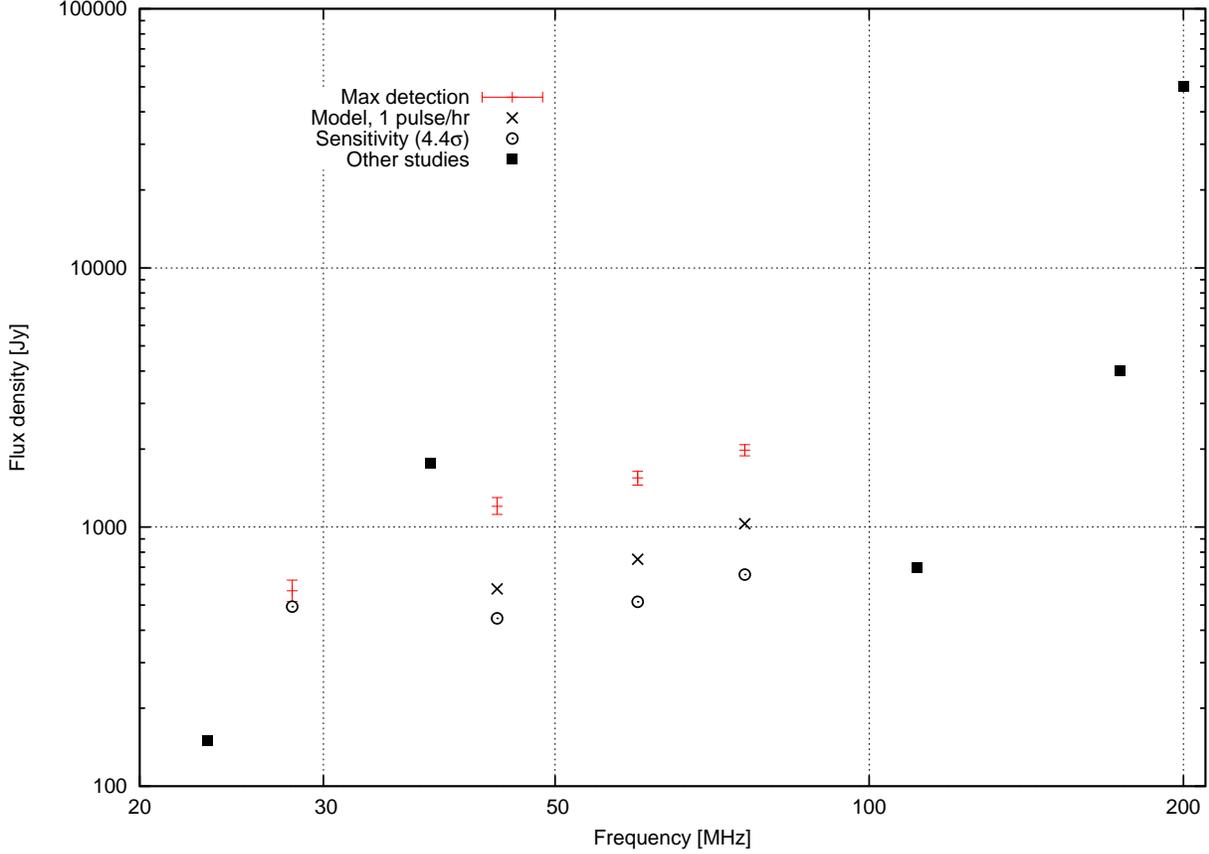}
%\vspace{1in}
\caption{
Red markers with error bars indicate the brightest pulses detected in the 28, 44, 60, and 76~MHz passbands in the present work.  Also shown are the effective pulse sensitivities ($4.4\sigma$) worked out in Section~\ref{sFC} and (for 44, 60, and 76~MHz only) flux densities at which a rate $>$ 1~pulse h$^{-1}$ is predicted, using Equations~\ref{eN76}--\ref{eN44}.  Included for comparison are points representing the brightest pulse in 12~h at 23~MHz reported by \cite{pop06}, the brightest pulse in 14~h at 38~MHz reported by \cite{des09}, the brightest pulse in 0.25~h at 111~MHz reported by \cite{pop06}, the brightest pulse in 1~h at 174~MHz reported by \cite{kar12}, and the brightest pulse reported in 3.5~h at 200~MHz by \cite{bha07}.
%Shown for comparison are the results reported in \citep{pop06}, \citep{des09}, and \citep{bha07} (23, 38, and 200~MHz respectively); these are shown both for the reported observing durations (12, x, and 3.5~h, respectively) as ``$\square$'' markers; and normalized to the expected value for 7~h of observation using the \citep{arg72} power law, as ``$\circ$'' markers.  are the fluxes for every pulse in this study which was detected at more than one frequency; values for a given pulse are connected by lines. 
\label{fFluxFreq} }
\end{figure}

Figure~\ref{fFluxFreq} shows the brightest pulses detected in each passband, the system sensitivity to Crab GPs (worked out at the end of Section~\ref{sFC}), and predicted flux densities for detection rates of 1~pulse/hour using Equations~\ref{eN76}--\ref{eN44}.  Because the brightest pulses represent a undersampled tail distribution, it is not reasonable to infer a spectral index from our data alone.  
%However comparison to previous work at 23, 38, and 200~MHz (also shown in Figure~\ref{fFluxFreq}) indicates our findings are consistent with 
%
\cite{bha07} suggest a power law of $\sim\nu^{+2.7}$ for flux density from a two-point fit using the 23 and 200~MHz data.  
%Our data combined with findings reported in \cite{kar12} may suggest a weaker dependence.  
Although the 174~MHz value from \cite{kar12} is for only 1~h, a model for the tail distribution of the PDF is also available; this model predicts that the brightest pulse in 10~h (for comparison with the present work) is 10.5 kJy.  This extrapolation combined with our data suggest that the dependence of flux density on frequency may be significantly less than $\nu^{+2.7}$.   

%Shown for comparison are the results reported in \citep{pop06}, \citep{des09}, and \citep{bha07} (23, 38, and 200~MHz respectively); these are shown both for the reported observing durations (12, x, and 3.5~h, respectively) and normalized to the expected value for 7~h of observation using the \citep{arg72} number density law. A best fit to the 7-hour maximum values at 23, 28, 44, 60, 76, and 200~MHz yields a power law with index $+x.x$ for flux density vs. frequency.\footnote{Excluded from the the fit are the 38~MHz value from \cite{des09} due to low peak S/N, and a 111~MHz value reported in \citep{pop06} due to short observation time (just 15 minutes), making a number density extrapolation unreasonable.} 

%Also shown in Figure~\ref{fFluxFreq} are the fluxes for every pulse in this study which was detected at more than one frequency; with values for each pulse connected by lines so as to identify the implied spectral index.  It is clear the data for individual pulses independently suggests a spectral index $\sim-x$.  This result is shown more clearly in Figure~\ref{fSIs}.  Note that the data are ...

%This power law, combined with the $A\nu^{-x}$ law derived for observations at frequencies greater than 430~MHz \citep{cor04}, suggests that the GP spectrum turns over at $\nu=x$~MHz.  This is consistent with early work suggesting a turnover in the pulsar spectrum $\sim200$~MHz \citep{man77}.

\subsection{\label{sPB}Pulse Broadening}

\begin{figure}
%\epsscale{.80}
\plotone{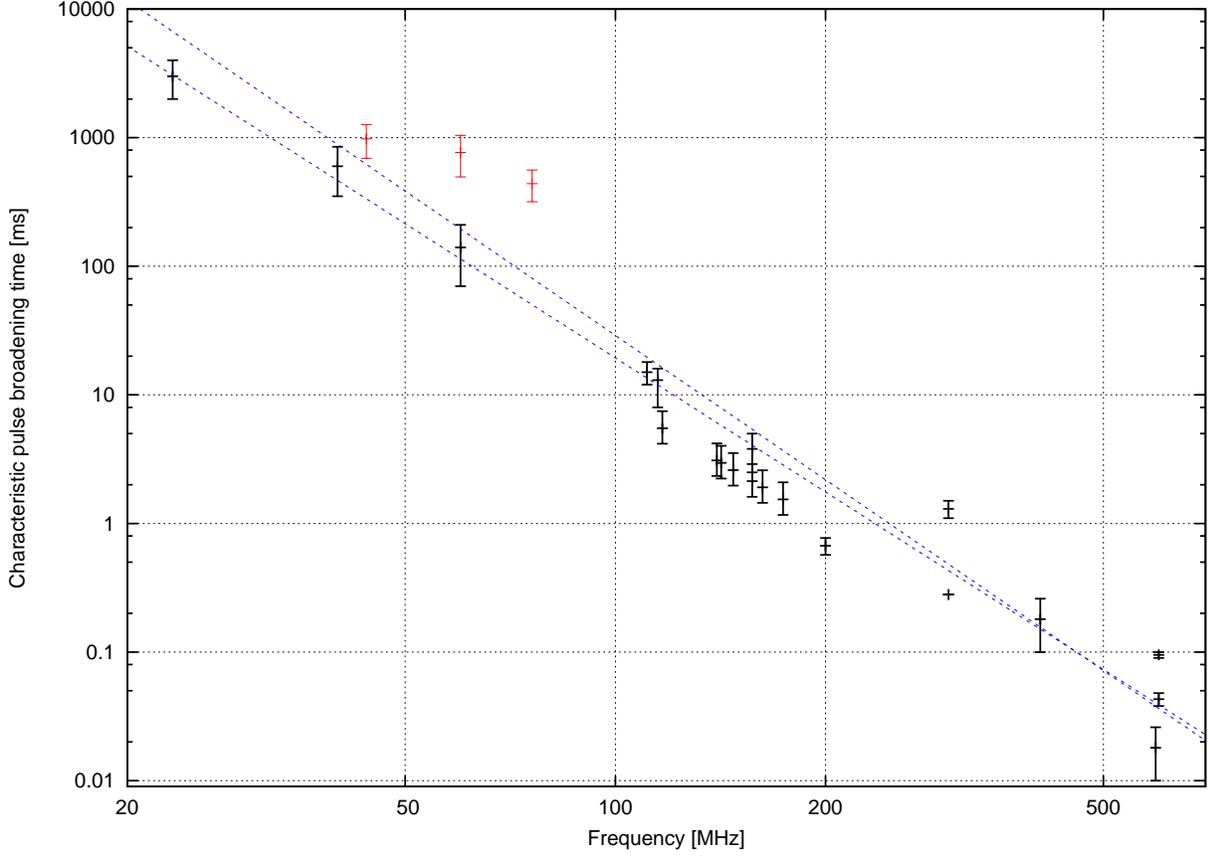}
%\vspace{1in}
\caption{
Characteristic pulse broadening time $\tau_d$ for GPs detected in the 44, 60, and 76~MHz passbands in the present work is shown in red. The error bars indicate the mean and $\pm1\sigma$ values.  
Also shown are measurements from previous studies listed in Table~\ref{tTaud}.
The lines are best-fit power laws using the mean values from all studies shown (upper line, $\alpha=-3.7$) and from all studies excluding the present work (lower line, $\alpha=-3.5$); see Table~\ref{tTaudPL}.
%Also included are results from previous work at 23 and 111~MHz \citep{pop06}, 116--174~MHz \citep{kar12}, 200~MHz \citep{bha07}, 300~MHz \citep{sal99}, and 600~MHz \citep{sal99} (higher value) and \citep{pop06} (lower value).  The lines are power law fits to the 44 \& 60~MHz mean values ($\alpha=-0.8$), the 60 \& 76~MHz mean values ($\alpha=-2.4$), and the 116--174~MHz mean values ($\alpha=-3.2$).
% Also shown is the best fit power law using the mean values at each frequency from this study combined with data from other studies.  The best fit implies a scaling of $\nu^{-3.6}$.
\label{fTD}  }
\end{figure}

Table~\ref{tTaud} shows statistics of $\tau_d$ and the pulse rise time parameter $\beta$ for all pulses considered in the study.  Figure~\ref{fTD} shows the statistics of $\tau_d$ in context with the analogous values for previous studies at other frequencies, which are also listed in Table~\ref{tTaud}.  As explained in Section~\ref{sPC}, the exponent $\alpha$ of the power law dependence of $\tau_d$ with frequency is of particular interest as it is related to the distribution of the scattering material.  Table~\ref{tTaudPL} shows $\alpha$ calculated using various combinations of the data represented in Figure~\ref{fTD} and Table~\ref{tTaud}.
Data from our study seem to indicate that (1) $\tau_d$ is significantly higher than might be expected by extrapolation from other studies, and (2) the dependence on frequency is significantly weaker than might be expected by extrapolation from other studies. 

With respect to the first finding, we have ruled out errors in assumed DM and residual dispersion (due to the use of incoherent dedispersion); we find that no combination of these mechanisms can account for a bias greater than about 35~ms in $\tau_d$, which is less than 8\% of the observed mean value for the highest-frequency passband.  The largest contribution to this bias is 28~ms due to error in the assumed DM.  This error is quantified as follows:  The Jodrell Bank Crab Pulsar ephemeris\footnote{http://www.jb.man.ac.uk/$\sim$pulsar/crab.html} \citep{lyn93} indicates that the actual DM for the months of October 2012 through January 2013 was consistently higher than the assumed DM, with the worst case error being $0.0644$~pc~cm$^{-3}$ larger than the assumed DM.
This corresponds to an error in dispersive delay of about 636~ms over the entire frequency range 20-82 MHz.  However we process this frequency span in separate 16 MHz passbands, and the worst case delay error across a passband is only about 109 ms (in the 44~MHz passband; it is worse in the 28~MHz passband but no pulse width measurements were obtained in that passband).  The bias subsequently induced in the estimate of $\tau_d$ is only about one-fourth of 109~ms (i.e., 28~ms), which was determined as follows:  We performed a simulation in which a pulse following the model of Equation~\ref{eMP} with $\beta=0.71$ and $\tau_d$=978~ms (mean values for pulses detected in the 44~MHz passband) was ``smeared'' by an error of 109 ms in dispersive delay, and then we used our model-fitting algorithm to determine the apparent values of $\beta$ and $\tau_d$.  We find that the apparent value of $\beta$ (as well as pulse amplitude) is not significantly affected and that the apparent value of $\tau_d$ is high by about 28~ms.  
%Subsequent experiments using this simulation confirm that the apparent $\tau_d$ is biased by about one-fourth of the dispersion delay error over the expected range of values of dispersion delay error, $\beta$, and $\tau_d$.  
The excess delay and resulting bias in all passbands is shown in Table~\ref{tFFT}.  
%Thus, error in estimates of $\tau_d$ introduced due to error in the assumed DM used for dedispersion are negligible.  

It is more likely that our data reflect actual changes in the distribution of ionized material within the nebula:  The Crab Pulsar is known to exhibit increases in $\tau_d$ by factors of 2--10 for intervals on the order of days to months, attributable to ionized clouds or filaments within the nebula crossing the line of sight \citep{ran73,lyn75,sal99,kuz08,sta11}.  This is consistent with the relatively high DM (noted above) in effect over the period of the observations.

Power laws in the form $\tau_d \propto \nu^{\alpha}$ implied by our data are included in Table~\ref{tTaudPL}.  
Recent studies (summarized in this table and in Table~\ref{tTaud}) have suggested that $\alpha$ lies in the range $-3.5$ to $-3.2$. This is not necessarily inconsistent with our data if we consider only the 76 and 60~MHz data.  However our 44~MHz data imply a flattening in the dependence with decreasing frequency which does not appear to be consistent with $\alpha \le -3.2$.      
The weak dependence on $\tau_d$ on frequency implied by our data, like the large values of $\tau_d$ implied by our data, might be interpreted as a perturbation due to inhomogeneity in the distribution of ionized material within the nebula.  \cite{cor01} have shown that inhomogeneities in the form of disks and filaments may lead to large variations in the dependence of $\tau_d$ on frequency, and that these effects may be particularly strong at low radio frequencies. 

Given that the studies identified in Table~\ref{tTaud} and Figure~\ref{fTD} represent widely-separated epochs of scattering behavior within the nebula, it may be too much to expect a simple dependence of $\tau_d$ on frequency to emerge from these data.
Thus a broader interpretation of our findings may be simply that the frequency dependence is complex: neither static nor well-described by a single power law extending to low radio frequencies.  
In any event, our results, like those of other recent studies, strongly suggest that the electron density distribution is not in the form of a Kolomogorov spectrum ($\alpha=-4.4$), and in fact $\alpha$ (to the extent that it is reasonable to characterize the frequency dependence of pulse broadening in this manner) is probably always significantly greater than $-4$. 

%The apparent flatness and frequency dependence of $\alpha$ in the present work is also difficult to explain.  First, it should be noted that the value of $\alpha$ obtained using the upper and lower $1\sigma$ values for 60 and 76~MHz respectively is $< -3.5$, so our findings do not rule out the possibility that the $\alpha=-3.5$ law holds to frequencies as low as $\sim60$~MHz.    The corresponding value obtained using the upper and lower $1\sigma$ values for 44 and 60~MHz respectively is only $-3.0$ and is thus harder to dismiss.  Also, the $\alpha=-0.8$ law for the 44--60~MHz data extrapolates to a value which significantly undershoots the 23~MHz data.  These findings might indicate a flattening in the frequency dependence of $\tau_d$ with decreasing frequency, perhaps to an asymptote ($\alpha \sim 0$).  If true, the physical interpretation is not obvious.  

\section{\label{sDisc}Concluding Remarks}

%\# Why 56.791 is OK; say what total variation is expected to be.  Characterize desensitization expected.  From Lyne website:  If you make use of these data in a publication, we request that you acknowledge the source of the information by referencing the paper: Lyne, A. G., Pritchard, R. S. \& Graham-Smith, F. 1993. MNRAS, 265, 1003, which describes the origin of the data, and by quoting the web address http://www.jb.man.ac.uk/~pulsar/crab.html.

We have demonstrated the ability of LWA1 to observe single Crab GPs over a frequency range of $\sim$4:1 at a rate of $\sim3$~h$^{-1}$, yielding information about pulse amplitude and pulse broadening that may be useful in gaining insight into the nature of the Crab GP emission mechanism as well as providing additional information about the electron density distribution within the nebula and along the line of sight.  The key to continued progress is observing a far greater number of pulses distributed over a timeframe significantly greater than the weeks-to-months-long scattering epochs.  
To this end, we are continuing our program of Crab GP observations with the goal of increasing the number of detections by an order of magnitude and then repeating the analysis presented above.
%, which we expect will provide improved insight into the static and time-varying nature of pulse broadening as well as pulse amplitude statistics that can be compared to higher-frequency studies.  
Also, we are preparing for simultaneous multiband observations of Crab GPs using LWA1 with 
%the Goldstone Deep Space Network 70~m radio telescope operating at 1.7~GHz, 
instruments operating at higher frequencies,
which may provide additional information on Crab GP behavior by allowing us to simultaneously observe emission on both sides of the turnover in the GP spectrum.


%% Included in this acknowledgments section are examples of the
%% AASTeX hypertext markup commands. Use \url without the optional [HREF]
%% argument when you want to print the url directly in the text. Otherwise,
%% use either \url or \anchor, with the HREF as the first argument and the
%% text to be printed in the second.

\acknowledgments

The authors acknowledge helpful discussions with
W.A. Coles, 
T.H. Hankins,
J.F. Helmboldt,
N.E. Kassim, 
W.A. Majid,
and
B.J. Rickett.
Some of the data presented in this paper were processed using the LWA1 User Computing Facility, which is a joint project of the University of New Mexico, Virginia Tech, and the Jet Propulsion Laboratory and in which J. Dowell was the primary contributor.
Basic research in astronomy at the Naval Research Laboratory is supported by 6.1 base funding. 
Construction of LWA1 was supported by the Office of Naval Research under Contract N00014-07-C-0147. 
Support for operations and continuing development of LWA1 is provided by the National Science
Foundation under grants AST-1139963 and AST-1139974 of the University Radio Observatories program.
The authors acknowledge the support of the National Radio Astronomy Observatory.

\clearpage

%% the file, uncomment the \plotone, \plottwo, and
%% If you need a layout that cannot be achieved with \plotone or
%% \plottwo, you can invoke the graphicx package directly with the
%% \includegraphics command or use \plotfiddle. For more information,
%% please see the tutorial on "Using Electronic Art with AASTeX" in the
%% documentation section at the AASTeX Web site,
%% http://www.journals.uchicago.edu/AAS/AASTeX.
%%
%% The examples below also include sample markup for submission of
%% supplemental electronic materials. As always, be sure to check
%% the instructions to authors for the journal you are submitting to
%% for specific submissions guidelines as they vary from

%% This example uses \plotone to include an EPS file scaled to
%% 80% of its natural size with \epsscale. Its caption
%% has been written to indicate that additional figure parts will be
%% available in the electronic journal.

\begin{table}
\begin{center}
\begin{tabular}{|l|rrrr|rr|}
\hline
MJD & 28~MHz & 44~MHz & 60~MHz & 76~MHz & 3~Passbands & 2~Passbands \\
\hline \hline
56221 & 0 & 10 & 10 & 9 & 7 & 2 \\
56223 & 0 &  0 &  2 & 3 & 0 & 3 \\
\hline
56275 & 0 &  1 &  2 & 2 & 1 & 1 \\
56276 & 0 &  4 &  3 & 3 & 3 & 0 \\
56277 & 0 &  0 &  1 & 1 & 0 & 1 \\
\hline
56285 & 0 &  2 &  2 & 2 & 2 & 0 \\
56286 & 0 &  5 &  5 & 5 & 5 & 0 \\
\hline
56289 & 0 &  1 &  1 & 0 & 0 & 1 \\
56291 & 1 &  3 &  3 & 3 & 3 & 0 \\
56292 & 0 &  2 &  4 & 3 & 1 & 3 \\
\hline \hline
TOTAL & 1 & 28 & 33 & 31 & 22 & 11 \\
$\ge 4.4\sigma$ & 1 & 21 & 32 & 29 &   &   \\
\hline
\end{tabular}
\end{center}
\caption{
Summary of daily observing sessions.  ``MJD'' is Modified Julian Date.  Each session was 1~hour in length.  Gaps of greater than 1 week are indicated by horizontal lines.  The center four columns are the number of pulses detected in each passband.  The rightmost two columns indicate the number of pulses appearing simultaneously in multiple passbands, respectively. The bottom row indicates the totals after excluding pulses having apparent peak S/N $< 4.4\sigma$ ($25$~ms, $\approx17$~MHz). 
\label{tObs}}
\end{table}

\begin{table}
\begin{center}
\begin{tabular}{|c||crrr|rr|}
\hline
Passband & Span       & $L$ & $\Delta\nu$ & $\Delta\tau$ & Excess Delay & Bias in $\tau_d$ \\
\hline
\hline
76~MHz   & 68--84~MHz &  16384     & 1196~Hz     &  836~$\mu$s    & $+20$~ms    &  $+5$~ms \\
60~MHz   & 52--68~MHz &  16384     & 1196~Hz     &  836~$\mu$s    & $+41$~ms    & $+10$~ms \\
44~MHz   & 36--52~MHz &  65536     &  299~Hz     & 3344~$\mu$s    & $+109$~ms   & $+28$~ms \\
28~MHz   & 20--36~MHz & 131072     &  150~Hz     & 6687~$\mu$s    & $+466$~ms   & $+117$~ms \\
\hline
\end{tabular}
\end{center}
\caption{
Time-frequency resolutions used for incoherent dedispersion.  Also shown is the additional dispersion delay across each passband due to worst-case error in the DM assumed for dedispersion (see Section~\ref{sDR}), and the associated bias in $\tau_d$ (see Section~\ref{sPB}).   
\label{tFFT}}
\end{table}

%\begin{figure}
%%\epsscale{.80}
%%\plotone{f1.eps}
%\vspace{1in}
%\caption{
%THIS FIGURE IS IN PREPARATION.  
%Time (bottom panel), frequency (top panel), and spectrogram (middle panel) representation of data collected over a x~s span during one of the observations.  The center frequency is 76~MHz.  Spectral boundaries of the 6~MHz ATSC channels are indicated, as well as the center frequencies of NTSC narrowband video and audio signals. \label{fRFI}}
%\end{figure}

%\begin{figure}
%\epsscale{.80}
%\plotone{f1.eps}
%\vspace{1in}
%\caption{Derived spectral index for each GP detected at more than one frequency, as a function of detection S/N.  In each case the S/N for the higher frequency is used.  The horizontal lines indicate the spectral indices implied from the analysis associated with Figure~\ref{fFluxFreq}, as well as spectral indices obtained from previous studies.
%\label{fSIs}  }
%\end{figure}

%\begin{table}
%\begin{center}
%\begin{tabular}{|ccc|}
%\hline
%Tuning & $\tau_d$ (ms) & $\beta$ \\
%\hline
%76~MHz & $439 \pm 122$ & $0.42 \pm 0.24$ \\
%60~MHz & $768 \pm 273$ & $0.56 \pm 0.16$ \\
%44~MHz & $978 \pm 287$ & $0.71 \pm 0.20$ \\
%\hline
%\end{tabular}
%\end{center}
%\caption{
%Pulse shape statistics: mean and standard deviation of $\tau_d$ and $\beta$ for all pulses in each tuning. 
%\label{tTaud}}
%\end{table}
%
%  76: mean=439.200000 ms, std=121.583990 ms
%  60: mean=767.620690 ms, std=272.700570 ms
%  44: mean=977.894737 ms, std=286.658507 ms
%  76-60-44 tau_d power law fit: -1.429848 12.357018
%  76-60 tau_d power law fit: -2.361960 16.313974
%  60-44 tau_d power law fit: -0.780598 9.839333
%mean beta 76/60/44/28:
%   0.42520   0.56172   0.71263   0.27000
%std  beta 76/60/44/28:
%   0.23737   0.16338   0.20044   0.00000

\begin{table}
\begin{center}
\begin{tabular}{|rcll|}
\hline
$\nu$ [MHz] & $\tau_d$ [ms] & Reference & Comment \\
\hline
23 & $3000 \pm 1000$ & \cite{pop06} & \\
40 & $600 \pm 250$ & \cite{kuz02} & \\
44 & $978 \pm 287$ & {\em This work} & $\beta=0.71 \pm 0.20$ \\
60 & $140 \pm 70$ & \cite{kuz02} & \\
60 & $768 \pm 273$ & {\em This work} & $\beta=0.56 \pm 0.16$ \\
76 & $439 \pm 122$ & {\em This work} & $\beta=0.42 \pm 0.24$ \\
%102 & $29$ & \cite{kuz96} & \\
111 & $15 \pm 3$ & \cite{pop06} & \\
115 & $13 \pm 5$ & \cite{sta70} & \\
157 & $3.8 \pm 1.3$ & \cite{sta70} & \\
174 & $1.5 \pm 0.4$ & \cite{kar12} & See caption \\ 
200 & $0.670 \pm 0.100$ & \cite{bha07} & \\
%200 & $0.395 \pm 0.100$ & \cite{bha07} & See caption\\
300 & $1.3 \pm 0.2$ & \cite{sal99} & See caption\\
300 & $0.28$ & \cite{sal99} & See caption\\
406 & $0.18 \pm 0.08$ & \cite{kuz02} & \\
594 & $0.018 \pm 0.008$ & \cite{kuz02} & \\
600 & $0.095 \pm 0.005$ & \cite{sal99} & \\
600 & $0.043 \pm 0.005$ & \cite{pop06} & \\
\hline
\end{tabular}
\end{center}
\caption{
Measurements of characteristic pulse broadening times shown in Figure~\ref{fTD}. 
\cite{kar12} reports values for 7 frequencies between 116.75 and 173.85 exhibiting $\approx\nu^{-3.2}$ dependence. 
%\cite{bha07} reports values using two different scattering models; the larger value is associated with $\tau_d$ as defined in this paper.
\cite{sal99} reports 300~MHz values for two widely-separated scattering epochs. 
\label{tTaud}}
\end{table}

\begin{table}
\begin{center}
\begin{tabular}{|ll|}
\hline
Data & $\alpha$ \\
\hline
All data  & $-3.7$  \\
All data except 76, 60, and 44~MHz values from this work & $-3.5$ \\
\cite{kar12} (117--174~MHz) & $-3.2 \pm 0.1$ \\
76 and 60~MHz values from this work (only) & $-2.4 \pm 2.9$ \\
76, 60, and 44~MHz values from this work (only) & $-1.4 \pm 1.1$ \\
\hline
\end{tabular}
\end{center}
\caption{
Power laws for the dependence of $\tau_d$ on frequency inferred from various combinations of the data in Table~\ref{tTaud}.  Shown is the exponent $\alpha$ for $\tau_d \propto \nu^{\alpha}$ as determined by log-linear least squares fit to the indicated data.
\label{tTaudPL}}
\end{table}

%% If you are not including electronic art with your submission, you may
%% mark up your captions using the \figcaption command. See the
%% No more than seven \figcaption commands are allowed per page,
%% so if you have more than seven captions, insert a \clearpage

%% Tables should be submitted one per page, so put a \clearpage before

%% Two options are available to the author for producing tables:  the
%% deluxetable environment provided by the AASTeX package or the LaTeX
%% table environment.  Use of deluxetable is preferred.
%%

%% The following command ends your manuscript. LaTeX will ignore any text


\begin{thebibliography}{}

\bibitem[Argyle \& Gower(1972)]{arg72}
  Argyle, E. \& Gower, J.F.R. 1972, \apj, 175, L89
\bibitem[Baars et al.(1977)]{bar77}
  Baars, J.W.M. et al. 1977, Astron.\ Astrophs., 61, 99
\bibitem[Bhat, Cordes \& Chatterjee(2003)]{bha03}
  Bhat, N.D.R., Cordes, J.M. \& Chatterjee, S. 2003, \apj, 584, 782 
\bibitem[Bhat et al.(2007)]{bha07} 
  Bhat, N.D.R. et al. 2007, \apj, 665, 618
\bibitem[Bietenholz et al.(1997)]{bie97}
  Bietenholz, M.F. et al. 1997, \apj, 490, 2991
\bibitem[Cordes et al.(2004)]{cor04}
  Cordes, J.M. et al. 2004 \apj, 612, 375
\bibitem[Cordes \& Lazio(2001)]{cor01}
  Cordes, J.M. \& Lazio, T.J.W. 2001, \apj, 549, 997
%Counselman, C. C. & Rankin, J. M., 1971. ApJ, 166, 513
\bibitem[Crossley et al.(2010)]{cro10}
  Crossley, J.H. et al. 2010, \apj, 722, 1908
\bibitem[Deshpande(2009)]{des09}
  Deshpande, K.B. 2009, M.S. Thesis, Virginia Tech. http://scholar.lib.vt.edu/theses/available/etd-11112009-104700/
\bibitem[Ellingson et al.(2013)]{ell13}
  Ellingson, S.W. et al. 2013, IEEE Trans. Ant. \& Prop., in press.  Preprint available: arXiv:1204.4816 [astro-ph.IM]
\bibitem[Hankins et al.(2003)]{han03}
  %Hankins, T.H., Kern J.S., Weatherall, J.C. \& Eilek, J.A. 2003, Nature, 422, 141
  Hankins, T.H. et al. 2003, Nature, 422, 141
\bibitem[Isaacman \& Rankin(1977)]{isa77}
  Isaacman, R. \& Rankin, J.M. 1977, \apj, 214, 214.
\bibitem[Karuppusamy, Stappers \& van Straten(2010)]{kar10}
  Karuppusamy, R., Stappers, B.W. \& van Straten, W. 2010, Astronomy \& Astrophysics, 515, A36
\bibitem[Karuppusamy, Stappers \& Lee(2012)]{kar12}
  Karuppusamy, R., Stappers, B.W. \& Lee, K.J. 2012, Astronomy \& Astrophysics, 538, A7
\bibitem[Kassim et al.(2007)]{k+07}
  Kassim, N.E. et al. 2007, \apjs, 172, 686
  %The Astrophysical Journal Supplement Series, 172:686Y719, 2007 October 2007
  %THE 74 MHz SYSTEM ON THE VERY LARGE ARRAY
\bibitem[Kuzmin et al.(2002)]{kuz02}
  Kuzmin, A. et al. 2002, Astron. Lett., 28, 251
\bibitem[Kuzmin et al.(2008)]{kuz08}
  Kuzmin, A. et al. 2008, A\&A, 483, 13
%\bibitem[Lundgren et al.(1995)]{lun95}
%  %Lundgren, S.C., Cordes, J.M., Ulmer, M., Matz, S.M., Lomatch, S., Foster, R.S. \& Hankins, T.H. 1995, \apj, 453, 433
%  Lundgren, S.C. et al. 1995, \apj, 453, 433
\bibitem[Lyne \& Thorne(1975)]{lyn75}
  Lyne, A.G. \& Thorne, D.F. 1975, MNRAS, 172, 97
\bibitem[Lyne, Pritchard \& Graham-Smith(1993)]{lyn93}
  Lyne, A.G., Pritchard, R.S. \& Graham-Smith, F. 1993, MNRAS, 265, 1003
\bibitem[Majid et al.(2011)]{maj11}
  Majid, W.A. et al. 2011, \apj, 741, 53
\bibitem[Manchester et al.(2005)]{man05}
  Manchester, R.N. et al. 2005, \aj, 129, 1993
%\bibitem[Manchester \& Taylor(1977)]{man77}
%  Manchester, R.N. \& Taylor, J.H. 1977, Pulsars (San Francisco: Freeman)
\bibitem[Mickaliger et al.(2012)]{mic12}
  Mickaliger, M.B. et al. 2012, ApJ, 760, 64
\bibitem[Moffett \& Hankins(1996)]{mof96}
  Moffett, D. A. \& Hankins, T.H. 1996, \apj, 468, 779
\bibitem[Parker(1968)]{par68}
  %Parker, E.A. 1968, Mon. Not. R. Astro. Soc., 138, 407
  Parker, E.A. 1968, MNRAS, 138, 407
\bibitem[Popov et al.(2006)]{pop06}
  Popov, M. V., et al. 2006, Astron. Rep., 50, 562
\bibitem[Rankin \& Counselman(1973)]{ran73}
  Rankin, J.M. \& Counselman III, C.C. 1973, \apj, 181, 875 
\bibitem[Roger, Costain \& Lacey(1969)]{rog69}
  Roger, R.S., Costain, C.H \&  Lacey, J.D. 1969, \aj, 74, 36
\bibitem[Roussel-Dupre, Jacobson \& Triplett(2001)]{rjt01}
  Roussel-Dupre, R.A., Jacobson, A.R. \& Triplett, L.A. 2001, Radio Sci., 36, 1615
  %Analysis of FORTE data to extract ionospheric parameters
  %Radio Science, Volume 36, Number 6, Pages1615-1630, November-December2001
\bibitem[Sallmen et al.(1999)]{sal99}
  %Sallmen, S., Backer, D.C., Hankins, T.H., Moffett, D. \& Lundgren, S. 1999, \apj, 517, 460
  Sallmen, S. et al. 1999, \apj, 517, 460
\bibitem[Staelin \& Reifenstein(1968)]{sta68}
  Staelin, D.H. \& Reifenstein, E.C., 1968, Science, 162, 1481
\bibitem[Staelin \& Sutton(1970)]{sta70}
  Staelin, D.H. \& Sutton, J.M. 1970, Nature, 226, 69
\bibitem[Stappers et al.(2011)]{sta11}
  Stappers, B.W. et al. 2011, A\&A, 530, A80
\bibitem[Stovall et al.(2013)]{sto13}
  Stovall, K. et al. 2013, in prep.
\bibitem[Taylor et al.(2012)]{tay12}
  Taylor, G. B. et al. 2012, J. Astro. Instr., 1. DOI: 10.1142/S2251171712500043. 
\bibitem[Williamson(1972)]{wil72}
  Williamson, I. P. 1972,  MNRAS, 157, 55
\end{thebibliography}
\end{document}